\documentclass[11pt]{article}
\usepackage[utf8]{inputenc}
\usepackage{enumitem}
\usepackage[skins,xparse,breakable]{tcolorbox}
\usepackage{amsmath,amsfonts,amssymb}
\usepackage{bbm}
\usepackage{dsfont}
\usepackage{amsthm}
\usepackage{hyperref}
\usepackage{fullpage}
\newtheorem{theorem}{Theorem}%[section]
\newtheorem{corollary}[theorem]{Corollary}
\newtheorem{lemma}[theorem]{Lemma}

\theoremstyle{definition}

\usepackage{bm}% bold math
\usepackage{thmtools}
\usepackage{thm-restate}
\usepackage[margin=0.95in]{geometry}
\usepackage{authblk}

\newcommand{\per}{\text{Per}}
\newcommand{\poly}{\text{poly}}

\usepackage{braket}

\title{On classical simulation algorithms for noisy Boson Sampling}
\date{\today}
\author[1]{Changhun Oh \thanks{changhun@uchicago.edu}}
\author[1]{Liang Jiang \thanks{liangjiang@uchicago.edu}}
\author[2]{Bill Fefferman \thanks{wjf@uchicago.edu}}

\affil[1]{\normalsize{Pritzker School of Molecular Engineering, University of Chicago, Chicago}}
\affil[2]{Department of Computer Science, University of Chicago, Chicago}

\begin{document}

\maketitle

\begin{abstract}

We present a classical algorithm that approximately samples from the output distribution of certain noisy Boson Sampling experiments.
This algorithm is inspired by a recent result of Aharonov, Gao, Landau, Liu and Vazirani and makes use of an observation originally due to Kalai and Kindler that the output probability of Boson Sampling experiments with a Gaussian noise model can be approximated by sparse low-degree polynomials.  This observation alone does not  suffice for classical sampling, because its marginal probabilities might not be approximated by sparse low-degree polynomials, and furthermore, the approximated probabilities might be negative. We solve this problem by employing the first quantization representation to give an algorithm for computing the marginal probabilities of these experiments.  

We prove that when the overall noise rate is constant, the algorithm runs in time quasi-polynomial in the number of input photons $N$ and accuracy.  When the overall noise rate scales as $1-x_1^\gamma$ for constant $x_1$ and $\gamma=\Omega(\log N)$, the running time becomes polynomial.

Furthermore, we study noisy Boson Sampling with practically relevant noise models such as partial distinguishability and photon loss.
We show that the same technique does not immediately apply in these settings, leaving open the possibility of a scalable demonstration of noisy quantum advantage for these noise models in certain parameter regimes.

%In particular, for partial distinguishability, the approximated distribution still requires an exponential cost to compute probabilities.
%For photon loss, the approximation technique requires choosing a sufficiently large degree to suppress the approximation error so that it can only work for a large photon loss regime.
%However, the large photon loss regime can already be simulated by other techniques using classical states, which suggests that there is no direct advantage from the low-degree approximation technique.
%It implies that the sparse low-degree polynomial approximation method requires a certain condition for the noise type.
\end{abstract}

\thispagestyle{empty}

\section{Introduction}

We have recently seen the first claims of experimental quantum advantage using both the random circuit sampling proposal implemented with superconducting qubits \cite{arute2019quantum, wu2021strong} as well as the Gaussian Boson Sampling proposal implemented in a linear optical architecture \cite{zhong2020quantum, zhong2021phase, madsen2022quantum}.  Such quantum advantage is a necessary step on the path toward building scalable, fault-tolerant quantum computers.  In addition quantum advantage is a fundamental milestone in its own right, where it can be interpreted as providing an experimental violation to the Extended Church-Turing thesis (see e.g., \cite{bernstein1993quantum, aaronson2011computational}).

With such an important milestone it is critical to analyze our evidence for believing that such experiments are classically intractable.  Here much is still unknown, and to improve this situation we must both bolster the classical hardness arguments as well as develop new classical simulation algorithms to challenge our assumptions.  In this work, inspired by a recent algorithm for simulating logarithmic depth noisy random quantum circuits due to Aharonov, Gao, Landau, Liu and Vazirani \cite{aharonov2022polynomial} and earlier work due to Gao and Duan \cite{gao2018efficient}, we develop a classical algorithm to approximately sample from the output distribution of certain noisy Boson Sampling experiments.  Much like the Aharonov et al. result, we do not expect that this algorithm is practical in its present form.  That is, it most likely will not ``spoof’’ present day Boson Sampling experiments in a reasonable amount of classical running time, due mainly to the inefficient scaling of the algorithm's running time with the noise rate.  Nonetheless we are able to prove that our algorithm works for a Gaussian noise model proposed in past work by Kalai and Kindler \cite{kalai2014gaussian}, which like depolarizing noise in the Aharonov et al. algorithm has the property that the noisy output distribution eventually converges to the uniform distribution.  We then discuss the prospects for extending this algorithm to other noise models, including photon loss and partial distinguishability.

\subsection{Putting recent simulation results in context}

After more than a decade of research in this area, there now is a body of work to support the classical intractability of these quantum advantage experiments.  This evidence comes primarily from complexity theoretic arguments proving that no efficient classical algorithm can simulate these experiments in the asymptotic regime as the system size increases under reasonable complexity theoretical assumptions (see e.g., \cite{di-terhal, Bremner459,aaronson2011computational, randomcircuits, ac, ag, noise, kondo, deshpande2021quantum, hangleiter2022computational}).

One potential challenge to these hardness arguments comes from uncorrected noise, which is perhaps the defining characteristic of near-term quantum computational experiments.  This noise degrades the quantum signal as the system size increases.  Consequently it is reasonable to expect that classical algorithms could potentially take advantage of this weakness to simulate noisy experiments at a sufficiently large system size.  While this has been an active subject of research with many results \cite{kalai2014gaussian, bremner2017achieving, gao2018efficient, renema2018classical, shchesnovich2019noise, garcia2019simulating, noh2020efficient, takahashi2021classically, qi2020regimes, oh2021classical, villalonga2021efficient}, we have arguably not yet seen a classical algorithm that simulates state-of-the-art quantum advantage experiments using a comparable amount of computational resources (see e.g., \cite{aaronson2022} for more discussion on this point).  Indeed, at the moment there is hope that near-term quantum advantage experiments operate in a “Goldilocks” regime in which the system size is large enough to be classically intractable to simulate, but not so large that uncorrected noise overwhelms the quantum signal\footnote{This “Goldilocks” regime is also important to enable classical verification techniques such as the cross-entropy benchmark, which currently requires exponential time on a classical computer.}.

Short of spoofing fixed size near-term experiments, one can ask if classical algorithms can efficiently simulate noisy quantum advantage experiments in the asymptotic limit as the system size scales.  Such a classical algorithm would rule out a fully scalable demonstration of quantum advantage with uncorrected noise.  Indeed, such a scalable demonstration would be of great interest, but until very recently was thought to be infeasible.  This pessimism was mainly due to two reasons.  The first major reason came from a foundational result due to Aharonov et al. from the late 1990’s \cite{aharonov1996limitations} showing that the total variation distance between the output distribution of a noisy quantum circuit with circuit depth $d$ and the uniform distribution is upper bounded by $2^{-O(d)}$ \footnote{Strictly speaking this upper bound applies to any quantum circuit that is subject to depolarizing noise with constant noise rate, although more recent results have clarified that it is widely applicable to a variety of reasonable noise models (see e.g., \cite{gao2018efficient, deshpande2022tight}).}.  This early result already rules out scalable quantum advantage for any depth that is super-logarithmic in the system size.  To make things worse, there is numerical evidence that the output distribution of most noisy \emph{random} quantum circuits converges to the uniform distribution at the even faster rate of $2^{-O(n\cdot d)}$ (see \cite{boixo2017fourier} and the corresponding discussion in \cite{noise}).  This rapid convergence would rule out scalable, noisy quantum advantage at \emph{any depth}.

The second major reason for pessimism came from a statistical property of the output distribution of random quantum experiments known as ``anticoncentration'', which is useful in the theoretical hardness analysis of these systems (see e.g., \cite{aaronson2011computational} for more discussion).  Anticoncentration is known to be a property of any ensemble of random quantum circuits that forms an approximate unitary two-design (see e.g., \cite{BrandaoHarrow,Eisert1,hangleiter2018anticoncentration}).  For $D$-dimensional local random quantum circuits with Haar random gates this property first arises at depth $n^{1/D}$ and this is believed to be optimal \cite{BrandaoHarrow,harrowmehraban}.  Consequently, if the spatial locality is constant, then combining this result together with the upper bound of Aharonov et al. \cite{aharonov1996limitations} we find that the noisy output distribution of such circuits is inverse superpolynomially close to the uniform distribution, which again rules out noisy, scalable quantum advantage in this regime.

However, in the last two years new results were proven which offered some brief hope that random quantum circuits might be able to achieve such a scalable noisy advantage at precisely logarithmic depth.  First the results of Dalzell et al. and Barak et al.  proved that random quantum circuits with Haar random two-qubit gates anticoncentrate at logarithmic depth\footnote{Strictly speaking this is proven for 1D and all-to-all connectivities, but is believed to hold for intermediate regimes such as a 2D grid.} \cite{dalzell2022random,barak-gao}.  Crucially, these papers directly analyze the anticoncentration property of the ensemble of circuits without relying on the approximate two-design property.  Moreover, these results are optimal, in the sense that sublogarithmic depth random quantum circuits with two-qubit Haar random gates are known \emph{not} to anticoncentrate \cite{dalzell2022random,deshpande2022tight}.

In addition, a result of Deshpande et al. proved that the total variation distance between the output distribution of most random quantum circuits and the uniform distribution is \emph{lower bounded} by a quantity that scales as $2^{-O(d)}$, matching the Aharonov et al. upper bound of $2^{-O(d)}$ \cite{deshpande2022tight}.  Putting these two results together gave rise to the (as it turns out, fleeting) hope that \emph{logarithmic depth} random quantum circuits with Haar random gates could offer a ``sweet-spot’’ regime in which the depth was both sufficient  to have anticoncentration yet shallow enough so that uncorrected noise does not overwhelm.

\subsection{The Aharonov et al. random circuit simulation algorithm}

This hope was very recently ruled out by a result of Aharonov et al. \cite{aharonov2022polynomial} which presents an efficient algorithm for approximately sampling from the output distribution of noisy random circuit ensembles that anticoncentrate, modulo the ``gate-set orthogonality'' constraint which is satisfied e.g., by two qubit Haar random gates.  This algorithm follows up on earlier work of Gao and Duan, which achieved the same accuracy in quasi-polynomial time \cite{gao2018efficient}.

Owing to the requirement of anticoncentration, these algorithms are useful for simulating random quantum circuits with depth that scales at least logarithmically in the system size \footnote{It still remains possible to prove hardness of sampling results for random quantum circuits with Haar random gates at sublogarithmic depths without needing anti-concentration, although it is likely that new ideas will be required. Additionally there exist ensembles of random circuits that anticoncentrate at \emph{constant} depths \cite{haferkamp2020closing} by using a distribution over gates that is very different from Haar random.  It remains unclear if the Aharonov et al. algorithm can be adapted to simulate such ensembles in the presence of noise.}.  
%These ensembles employ gates that are not Haar random.  caveat is 
 %the possibility of the existence of random circuits having anti-concentration in sublogarithmic depth. Indeed, there exists a Hamiltonian with random inputs that achieves the anti-concentration property in a constant time .}}. 
In particular at logarithmic depth the earlier Aharonov et al. result implies that sampling from the uniform distribution  achieves total variation distance $1/2^{O(d)}=1/poly(n)$ \cite{aharonov1996limitations}.  
However, approximating the noisy output distribution by the uniform sampler cannot reduce the total variation distance by increasing the running time because the approximate sampler is fixed.
By contrast this new result is stronger and gives a classical algorithm that can achieve \emph{any} total variation distance parameter $\epsilon$ with a running time that scales as $\poly(1/\epsilon)$. 

The key observation behind this algorithm is that the output (or marginal) probabilities of noisy random circuits with a constant rate of depolarizing noise per gate can be expressed as the sum of polynomially many dominant Fourier coefficients with exponentially many other Fourier coefficients that are highly suppressed due to the noise.
In other words, the output probability of noisy random circuits can be approximately represented by sparse Fourier coefficients with a small error occurring by discarding other Fourier coefficients.
Using sparsity of the Fourier coefficients involved in the output (or marginal) probabilities, one can efficiently approximate the output (marginal) probabilities, which enables us to sample from the distribution.
We emphasize that it is crucial that {\it any} output probability of a given circuit has to be described by the {\it same} polynomially many Fourier coefficients to guarantee that all the marginals can also be efficiently computed.
The latter is not obvious because the marginal probabilities can be the sum of exponentially many probabilities, which may eventually require an exponential number of Fourier coefficients even though each probability has a sparse Fourier description.
In addition, since the approximated distribution can be a quasi-probability distribution, i.e., it can be negative, it was crucial to exploit a technique proposed in \cite{bremner2017achieving}, which enables us to approximately sample from a proper probability distribution when the quasi-probability distribution is sufficiently close to the noisy probability distribution.

\subsection{Noisy Boson Sampling} \label{sec:noisy_BS}
Let us turn our attention to Boson Sampling \cite{aaronson2011computational}, which is our main focus in the present work.
The main question of the present work is whether the same type of Aharonov et al. classical algorithm \cite{aharonov2022polynomial} works to simulate noisy Boson Sampling.
Interestingly, even before studies on the sparsity of Fourier coefficients in noisy random circuit sampling \cite{gao2018efficient, aharonov2022polynomial}, Kalai and Kindler already pointed out that low-degree polynomials can approximate the output probability of noisy Boson Sampling with a particular choice of noise type, which transforms a given linear-optical circuit $U\to \sqrt{x}U+\sqrt{1-x}Y$, where $Y$ is a random Gaussian matrix and $1-x$ is the noise rate.
To avoid  confusion we emphasize that $1-x$ is the noise rate not $x$, which is the case in \cite{aharonov2022polynomial}.
After Kalai and Kindler's analysis on a mathematically appealing noise model, several subsequent works studied more physical noise types such as partial distinguishability using similar techniques \cite{renema2018efficient, renema2018classical, shchesnovich2019noise, moylett2019classically}.
However, the previous works did not provide a classical sampler to exploit the low-degree polynomial approximation (See Sec.~\ref{sec:previous_BS} for more details).

In this work, we present a classical algorithm that approximately simulates noisy Boson Sampling with noise studied in ~\cite{kalai2014gaussian} using sparsity of low-degree polynomials and the method in \cite{bremner2017achieving}.
In particular, assuming Haar-random linear-optical circuits (instead of anticoncentration), the classical algorithm's running time is given by quasi-polynomial in the system size and accuracy for an overall constant noise level $1-x\in (0,1]$:
\begin{theorem}\label{thm1}
    Consider an $M$-mode Fock-state Boson Sampling with $N$ single photons and a linear-optical circuit with a global Haar-random unitary with $M=\omega(N^5)$.
    If there is an overall constant circuit noise, we can classically simulate collision-free outcomes of the noisy Boson Sampling with running time $N^{O(\log N, \log \epsilon^{-1},\log \delta^{-1})}$ within total variation distance $\epsilon$ for $1-\delta$ portion of Haar-random unitary matrices.
\end{theorem}
The main reason that the running time is quasi-polynomial is that the noise rate is assumed constant for the entire circuit instead a constant level of noise per gate as in \cite{aharonov2022polynomial}, where noise scales with the system size.
To introduce a similar effect, we now consider the case where the total noise rate scales as $1-x_1^\gamma$ with $\gamma={\Omega(\log N)}$ and a constant $x_1\in[0,1)$ and show for this case that the running time becomes polynomial:
\begin{corollary}\label{thm2}
    Consider an $M$-mode Fock-state Boson Sampling with $N$ single photons and a linear-optical circuit with a global Haar-random unitary with $M=\omega(N^5)$.
    If there is an overall circuit noise $1-x_1^\gamma$ with a constant $x_1\in[0,1)$ and $\gamma=\Omega(\log N)$, we can classically simulate collision-free outcomes of the noisy Boson Sampling with running time ${\poly(N, \epsilon^{-1},\delta^{-1})}$ within total variation distance $\epsilon$ for $1-\delta$ portion of Haar-random unitary matrices.
\end{corollary}
Note that whereas \cite{aharonov2022polynomial} introduces noise for each gate, but also requires anticoncentration, we introduce the noise for the entire circuit at once with global Haar-random circuits but do not explicitly require anticoncentration. It remains open to generalize our result as the setting in \cite{aharonov2022polynomial}.

The key idea to channel the sparse low-degree polynomial approximation from \cite{kalai2014gaussian} to sampling is to employ the first quantization representation of Boson Sampling.
We show that the marginals of approximated quasi-probability distribution for the first quantization representation can also be efficiently computed by sparse polynomials, and consequently the technique from \cite{bremner2017achieving} can be applied for sampling.
Thus, it closes the gap between the approximate computation of probability and sampling for circuit noise.
Intriguingly, applying the same sparsity technique to physical noise models such as partial distinguishability and photon loss hits barriers to finding a corresponding classical sampler.
First, for partial distinguishability noise, the barrier is that even after introducing noise and approximating the probability with similar polynomials, computing the output probability distribution still costs an exponential time.
Thus, a naive approach does not successfully reduce the complexity by exploiting the noise.
%It can be predicted by observing that the probability of the fixed point of partial distinguishability noise, fully distinguishable Boson Sampling, is still written as a permanent of a positive matrix.
Second, for photon loss, the barrier is that we need to choose a large degree to suppress the approximation error, which implies that the algorithm might work only for a large photon-loss regime.
However, the large photon-loss regime can already be classically simulated because lossy single-photon states are already sufficiently close to classical states (much like the convergence of the output probability distribution to uniform at superlogarithmic depth for qubit cases \cite{aharonov1996limitations}) \cite{oszmaniec2018classical, garcia2019simulating, qi2020regimes}.
Thus, the sparsity technique does not provide any benefits over the existing methods.

Our analysis of three different types of noise clearly reveals that the different behavior of output distributions against different noise types poses difficulties in the generalization of the same technique for more general noise models.
Interestingly, both the output distribution of random circuits with depolarizing noise and that of Boson Sampling with circuit noise converge to the uniform distribution, while those of Boson Sampling with partial distinguishability and photon loss do not.
This might indicate that the current technique implicitly relies on a certain property of the noise model, which is related to convergence to the uniform distribution, and that different noise models might require an additional technique or perhaps even lead to a scalable demonstration of noisy quantum advantage.  We stress, however, that we do not prove such a formal connection to the uniform distribution in this work, but leave this as an intriguing open direction for future research. 

\subsection{Relation to previous results on Boson Sampling}\label{sec:previous_BS}
As mentioned in the previous section, the low-degree polynomial approximation techniques for noisy Boson Sampling have been discussed even before \cite{gao2018efficient, aharonov2022polynomial}.
More specifically, Kalai and Kindler showed that the output probabilities of noisy Boson Sampling can be approximated by sparse low-degree polynomials under the assumption of Haar-randomness of the linear optical circuit matrix (this seems analogous to the anticoncentration requirement of Aharonov et al. \cite{aharonov2022polynomial}) \cite{kalai2014gaussian}.
Nevertheless, it is not obvious how to approximately {\it sample} from the output distribution described by the sparse low-degree polynomials because the approximated distribution might not be a proper probability distribution and it is not guaranteed that its marginal probabilities can also be described by sparse polynomials.
The latter is because it has to be shown that {\it any} probabilities can be described by the same sparse low-degree polynomials.
Our contribution is to channel the low-degree polynomial approximation to a classical sampling algorithm using the first quantization method and marginal-based sampler.

Several subsequent works studied more physical noise types such as partial distinguishability \cite{renema2018efficient, renema2018classical, shchesnovich2019noise, moylett2019classically} while their approaches also encounter the same obstacles to finding a classical sampler \footnote{While \cite{shchesnovich2019noise} claimed that there is an efficient classical sampler, this was not completely proved to the best of our knowledge.}.
In particular, \cite{renema2018efficient} observed that the output probability of partial distinguishable Boson Sampling can be approximated by low-degree polynomials, which guarantees that the total variation distance can be made small by choosing an appropriate degree.
It was also claimed that each polynomial can be efficiently approximated (not exactly computed, unlike \cite{kalai2014gaussian, aharonov2022polynomial}).
Nevertheless, it did not analyze the effect of the approximation of polynomials and did not provide a provable classical sampler; instead, it considered the Metropolis algorithm, which is heuristic \cite{neville2017classical}.
Thus, they did not provide a provable classical sampler for partial distinguishable Boson Sampling.
%A caveat is that the approximated distribution is not guaranteed to be a proper probability distribution and that the Metropolis algorithm is heuristic.
We show that indeed it is not immediately straightforward to construct a classical sampler that exploits the low-degree polynomial approximation for partial distinguishable noise.

Finally, there have been extensive studies on the effect of photon loss on Boson Sampling \cite{aaronson2016bosonsampling, oszmaniec2018classical, renema2018classical, garcia2019simulating, shchesnovich2019noise, qi2020regimes, oh2021classical}, while a similar technique has not been considered \footnote{\cite{shchesnovich2019noise} has considered the combined effect of loss and dark count with assuming that the total photon number is preserved by dark count effect, which is not satisfied solely by photon loss.}.
Our analysis shows that the previous techniques that approximate lossy single photons by classical states provide a better approximation error than a naive approach using the low-degree polynomial approximation.

\subsection{Concluding remarks}
We finally remark on several points that were not addressed in the present work and open questions.
\begin{itemize}
%    \item {\bf Polynomial-time classical algorithm.} An obvious open question from our main results is to improve our quasi-polynomial time algorithm further to a polynomial time algorithm.
    %Circumventing the barriers
    \item {\bf Efficient classical algorithms for physical noise models.} As we claimed, the low-degree polynomial approximation does not immediately lead to an efficient classical sampler for partial distinguishability and photon loss, which are the most crucial noise models in practice \cite{zhong2020quantum, zhong2021phase, madsen2022quantum}. It remains an open question to improve the technique to find an efficient classical sampler for those noise models.
    For photon loss case in particular, when the output photon number scales as $\Theta(\sqrt{N})$, the total variation distance of the classical algorithms in \cite{oszmaniec2018classical, garcia2019simulating,qi2020regimes} to the lossy output probability distribution is fixed as a constant, and it cannot be reduced by increasing the running time of the algorithms.
    Finding a classical algorithm that can efficiently reduce the approximation error as \cite{aharonov2022polynomial} and our result for Gaussian noise is another open question.
    \item {\bf Lifting the assumption of global Haar-randomness.} In the present work, we have assumed that the linear-optical circuits are constructed to be global Haar-random\footnote{Unlike random circuit sampling using qubits, the dimension of the unitary matrix for global Haar-random is polynomial in the system size. Thus, it is not an unrealistic assumption in practice (see e.g., \cite{russell2017direct}).}, which is a standard assumption for the hardness of Boson Sampling \cite{aaronson2011computational}.
    On the other hand, the recent Boson Sampling experiments have not implemented global Haar-random circuits \cite{zhong2020quantum, zhong2021phase, madsen2022quantum, oh2022classical}.
    Also, the recent result for random circuits \cite{aharonov2022polynomial} assumed anticoncentration with consideration of depth and noise effect per gate.
    Extending our results further with a less stringent assumption is another future work, such as replacing the global Haar-random assumption with anticoncentration.
    Note that whereas random circuits in \cite{aharonov2022polynomial} with gate-set orthogonality enjoy the symmetry between different outcomes when averaged over ensembles, Boson Sampling outcomes generally do not have such an apparent symmetry, which hinders us from analyzing the upper bound of total variation distance except for the global Haar-random case.
    \item {\bf Anticoncentration of Boson Sampling.} Unlike random circuit sampling, we have less understanding of anticoncentration in Boson Sampling such as how much circuit depth is required to attain anticoncentration property with what kinds of an ensemble of linear-optical circuits. Even whether the anticoncentration property is achieved with global-Haar random remains a conjecture to the best of our knowledge \cite{aaronson2011computational} despite interesting recent progress (see e.g., \cite{nezami2021permanent}).
    \item {\bf Gaussian Boson Sampling.} We have considered Fock-state Boson Sampling only while the quantum advantage experiments employed Gaussian Boson Sampling \cite{hamilton2017gaussian, deshpande2021quantum}, which is a variant of Fock-state Boson Sampling.
    While we expect a similar result to hold, we leave it as an open question.
    \item {\bf Practical consideration.} As emphasized before, the proposed algorithm assumes an asymptotic regime of noisy Boson Sampling and we do not expect the algorithm to spoof finite-size near-term experiments. Specifically, for a small noise rate $x_1\approx 1$, the degree of the polynomial of the running time is given by $1/\log(1/x_1)\approx 1/(1-x_1)$, which makes the algorithm impractical. In fact, the recent result in \cite{aharonov2022polynomial} observed the same issue, i.e., the degree of the polynomial is a large constant $1/\gamma$, where $\gamma$ is the noise rate per gate in their notation. An interesting future work is to improve the algorithm to be applicable to finite-size Boson Sampling.
\end{itemize}

\section{Fock-state Boson Sampling in first quantization}
Let us consider the standard Fock-state Boson Sampling \cite{aaronson2011computational}.
The basic setup is to prepare $N$ single photons and to inject the photons into an $M$-mode linear-optical circuit $\hat{U}$, characterized by an $M\times M$ unitary matrix, where $M=\poly(N)$.
We then measure the number of photons for each output mode, which gives rise to a measurement outcome $\bm{m}\in\mathbb{Z}_{\geq 0}^M$ with $\sum_{i=1}^M m_i=N$, where $m_i$ represents the number of photons at the $i$th output mode.
We now describe the dynamics by introducing the first quantization representation, which enables us to analyze marginal distributions later easily.
First, we write the input state as
\begin{align}
    \frac{1}{\sqrt{N!}}\sum_{\sigma\in\mathcal{S}_N}|\sigma(1),\dots,\sigma(N)\rangle,
\end{align}
where $\mathcal{S}_N$ represents the permutation group for $N$ elements, which accounts for the symmetrization of $N$ photons due to bosons' indistinguishability nature.
Thus, the density matrix of the input state is written as
\begin{align}
    \frac{1}{N!}\sum_{\sigma,\rho\in\mathcal{S}_N}|\sigma(1),\dots,\sigma(N)\rangle
    \langle\rho(1),\dots,\rho(N)|.
\end{align}
After applying beam splitter network $\hat{U}$, we obtain the output state
\begin{align}
    \frac{1}{N!}\sum_{\sigma,\rho\in\mathcal{S}_N} \hat{U}^{\otimes N}|\sigma(1),\dots,\sigma(N)\rangle
    \langle\rho(1),\dots,\rho(N)|\hat{U}^{\dagger\otimes N},
\end{align}
where the linear-optical operation, characterized by an $M\times M$ unitary matrix $U$, transforms the state as
\begin{align}
    \hat{U}|i\rangle= \sum_{j=1}^M U_{ij}|j\rangle.
\end{align}
Finally, we measure in each photon's position $\bm{r}\in \mathbb{Z}_{\geq 0}^N$, whose probability is written as
\begin{align}
    p(\bm{r})
    &=\frac{1}{N!}\sum_{\sigma,\rho\in\mathcal{S}_N} \langle \bm{r}|\hat{U}^{\otimes N}|\sigma(1),\dots,\sigma(N)\rangle
    \langle\rho(1),\dots,\rho(N)|\hat{U}^{\dagger\otimes N}|\bm{r}\rangle
    =\frac{1}{N!}\sum_{\sigma,\rho\in\mathcal{S}_N} \left(\prod_{i=1}^N U_{\sigma(i),r_i}U^*_{\rho(i),r_i}\right).
\end{align}
Especially for collision-free outcomes $\bm{r}$, i.e. at most a single photon clicks for each output mode (equivalently all $r_i$'s are distinct), the probability reduces to
\begin{align}
    p(\bm{r})=\frac{|\per U_{N,\bm{r}}|^2}{N!},
\end{align}
where $U_{N,\bm{r}}$ is the $N\times N$ submatrix of a unitary matrix $U$ obtained by selecting the first $N$ rows, which accounts for the input photons, and $\bm{r}$'s columns.
%We note that the output probability is the same as the one in the first type of Clifford-Clifford algorithm \cite{clifford2018classical}.

We first clarify the notation of outcomes $\bm{m}$, $\bm{r}$, and $\bm{z}$ and their relations, the latter of which will be defined now.
First, we will define $\bm{z}\in\mathbb{Z}_{\geq0}^N$ as the ordered vector of $\bm{r}$ in the nondecreasing order, i.e., $z_1\leq z_2\leq \dots \leq z_N$.
Notice that different $\bm{r}$'s may reduce to the same vector $\bm{z}$, which is because we cannot distinguish which input photons correspond to which output photons in principle due to the indistinguishability.
Because of the symmetry, the different $\bm{r}$'s that correspond to the same $\bm{z}$ have the same probability.
The photon number vector $\bm{m}$'s elements $m_i$'s can be obtained by counting the number of $i$'s in $\bm{z}$. 
Hence, we can write the probability by abusing the notation of $p(\cdot)$
\begin{align}
    p(\bm{z})=\sum_{\sigma\in\mathcal{S}_N}p(\sigma(\bm{r}))=|\per U_{N,\bm{r}}|^2.
\end{align}
We will often abuse the notation of the probability $p(\bm{z})$, $p(\bm{r})$ and $p(\bm{m})$, which can be uniquely identified by using different arguments $\bm{z}, \bm{r}$ and $\bm{m}$.

Especially when another distribution $q(\bm{z})$ has the same property, namely $q(\bm{z})=\sum_{\sigma\in\mathcal{S}_N}q(\sigma(\bm{r}))$, the total variation distance between $p(\bm{z})$ and $q(\bm{z})$ with ordered outcomes and that between $p(\bm{r})$ and $q(\bm{r})$ with unordered outcomes are equal:
\begin{align}\label{eq:error_relation}
    \|p(\bm{z})-q(\bm{z})\|_1
    =\|p(\bm{r})-q(\bm{r})\|_1.
\end{align}
The property will play an important role for approximate sampling.

We will focus on the (strong) collision-free regime $M=\omega(N^5)$, where an $N\times N$ submatrix of an $M\times M$ Haar-random unitary matrix $U$ can be approximated by complex random Gaussian matrix $Z$ such that $(U_{N,\bm{z}})_{ij}\approx Z_{ij}/\sqrt{M}$ with $Z_{ij}\propto \mathcal{N}(0,1)$ with scaling factor $1/\sqrt{M}$ \cite{aaronson2011computational}.
Also, we will focus on simulating the probability distribution over collision-free outcomes, which suggests that $r_i\neq r_j$, or equivalently $z_i\neq z_j$, for all $i\neq j\in [N]$, or $m_i\in \{0,1\}$ for all $i\in[M]$.
Since we do not aim to simulate collision outcomes, we will set all the collision outcomes to be $c$, i.e., we treat them as the same outcome $c$.

\section{Low-degree polynomial approximation with circuit noise}
\subsection{Noise sensitivity and low-degree polynomial approximation}
Let us consider the effect of noise on Boson Sampling output probability distributions.
The first type of Gaussian noise we consider is the noise on circuit unitary $U$.
Although this type of noise might not be physically or experimentally relevant, it provides profound insights into the noise sensitivity of the output probability of Boson Sampling.
More specifically, the introduced noise changes a unitary matrix as $U\to \sqrt{x}U+\sqrt{1-x} Y$, where $Y$ is an $M\times M$ complex random Gaussian matrix, $x\in [0,1]$, and $1-x$ is the noise rate \cite{kalai2014gaussian}.
We remark that the definition of the Gaussian noise seems to be unphysical in the sense that for a single instance $Y$, $\sqrt{x}U+\sqrt{1-x}Y$ is not necessarily unitary and, furthermore, its spectral norm can be larger than 1.
However, we show that the noisy output probability distribution is a proper probability distribution (see Appendix~\ref{app:kalai}).

In this section, we will recall the result from~\cite{kalai2014gaussian} that a Boson Sampling probability distribution under this type of noise can be approximated in total variation distance by low-degree polynomials.
To this end, let us consider an output probability
\begin{align}
    p(\bm{z})
    =|\per U_{N,\bm{z}}|^2
    =\frac{|\per(Z)|^2}{M^{N}},
\end{align}
where $Z\equiv \sqrt{M} U_{N,\bm{z}}$ is the rescaled $N\times N$ submatrix of unitary $U$ corresponding to the outcome $\bm{z}$.
We used the fact that a submatrix of a large Haar-random unitary matrix $(M=\omega(N^5))$ can be approximated by a complex random Gaussian matrix whose elements follow complex normal distribution $\mathcal{N}(0, 1)$ \cite{aaronson2011computational}.
Following ~\cite{kalai2014gaussian}, we can expand the absolute-squared permanent as the sum of orthogonal polynomials:
\begin{align}
    |\per (Z)|^2=\sum_{k=0}^N f^{=2(N-k)},
\end{align}
where the degree $2(N-k)$ polynomials $f^{=2(N-k)}$ satisfy the following orthogonal relations
\begin{align}
    \mathbb{E}_Z[|f^{=2(N-k)}|^2]=(N!)^2, ~~~ 
    \mathbb{E}_Z[f^{=2(N-k_1)}f^{=2(N-k_2)*}]=0, ~~\text{for}~~k_1\neq k_2.
\end{align}
Here, the average $\mathbb{E}_Z[\cdot]$ is taken over complex Gaussian random matrices $Z$.
To see this, let us expand the absolute-squared permanent of a complex random Gaussian matrix as
\begin{align}
    |\text{Per}(Z)|^2
    &=\sum_{\sigma,\rho\in\mathcal{S}_N}\prod_{i=1}^N z_{\sigma(i),i}z^*_{\rho(i),i} \\ 
    &=\sum_{\sigma,\rho\in\mathcal{S}_N}\prod_{i\in T}(z_{\sigma(i),i}z^*_{\sigma(i),i})\prod_{i\in T^c}(z_{\sigma(i),i}z^*_{\rho(i),i}) \\ 
    &=\sum_{\sigma,\rho\in\mathcal{S}_N}\prod_{i\in T}(1+h_2(z_{\sigma(i),i}))\prod_{i\in T^c}(z_{\sigma(i),i}z^*_{\rho(i),i}) \\ 
    &=\sum_{\sigma,\rho\in\mathcal{S}_N}\sum_{R \subset T}\left[\prod_{i\in T\setminus R}h_2(z_{\sigma(i),i})\prod_{i\in T^c}z_{\sigma(i),i}z^*_{\rho(i),i}\right] \label{eq:per_sq},
\end{align}
where we defined $T\subset[N]$ as the set of indices such that $\sigma(i)=\rho(i)$ for given permutations $\sigma$ and $\rho$ and $h_2(z)\equiv zz^*-1$.
An important fact is that $\{1, z, z^*$, $h_2(z)\}$ forms an orthogonal basis, i.e., $\mathbb{E}_Z[f_1 f_2^*]=0$ if $f_1$ and $f_2$ are different functions out of the basis, and they are eigenvectors of the noise operator $T_x[f](z)\equiv \mathbb{E}_y[f(\sqrt{x}z+\sqrt{1-x}y)]$ with $y$ being the complex random Gaussian noise $\mathcal{N}(0, 1)$, namely,
\begin{align} \label{eq:noise}
    1\to 1, ~~~z\to \sqrt{x} z, ~~~z^*\to \sqrt{x} z^*, ~~~\text{and}~~~ h_2(z)\to x h_2(z).
\end{align}
Here, we assign a degree for each by adding 1 for $z$ or $z^*$ and 2 for $h_2$.
Thus, the degree of the term in the parenthesis in Eq.~\eqref{eq:per_sq} is $2(|T|-|R|)+2(N-|T|)=2(N-|R|)$.
We further partition these terms according to the image $R'$ of $R$ under $\sigma$ and $\rho$.
Thus, we denote by $\sigma'$ and $\rho'$ the restriction of $\sigma$ and $\rho$ on the complement of $R$, namely these are one-to-one functions from $R^c$ and $[N]\setminus R'$.
Let $S(\sigma',\rho')\subset R^c$ be the set of indices on which they agree.
Using some algebra, one can show that the degree $2(N-k)$ part is given by
\begin{align}
    f^{=2(N-k)}
    &=\sum_{\substack{R,R'\subset [N]:\\ |R|,|R'|=k}}\sum_{\substack{\sigma \in \mathcal{S}_k:\\ R\to R'}}\sum_{\substack{\sigma',\rho'\in \mathcal{S}_{N-k}: \\ R^c\to R'^c}}\prod_{i\in S(\sigma',\rho')}h_2(z_{\sigma'(i),i})\prod_{i\in R^c\setminus S(\sigma',\rho')} z_{\sigma'(i),i}z^*_{\rho'(i),i} \\
    &=\sum_{\substack{R,R'\subset [N]:\\ |R|,|R'|=k}}k!\sum_{\substack{\sigma',\rho'\in \mathcal{S}_{N-k}: \\ R^c\to R'^c}}\prod_{i\in S(\sigma',\rho')}h_2(z_{\sigma'(i),i})\prod_{i\in R^c\setminus S(\sigma',\rho')} z_{\sigma'(i),i}z^*_{\rho'(i),i}.
\end{align}
Hence, we can rewrite the absolute-squared permanent as
\begin{align}
    |\per (Z)|^2=\sum_{k=0}^N f^{=2(N-k)},
\end{align}
as desired.

Let us introduce the noise.
As shown from Eq.~\eqref{eq:noise}, the noise operator $T_x[f](z)$ introduces additional prefactor $x^{N-k}$ for each $2(N-k)$-degree polynomial, i.e., 
\begin{align}
    f^{=2(N-k)}\to x^{N-k}f^{=2(N-k)}.
\end{align}
Hence, the noisy output probability becomes
\begin{align}
    \tilde{p}(\bm{z})
    =\frac{1}{M^{N}}\sum_{k=0}^N x^{N-k} f^{=2(N-k)}.
\end{align}
Also, the following relations can be easily checked from the orthogonality of basic elements~\cite{kalai2014gaussian}:
\begin{align}\label{eq:ortho}
    \mathbb{E}_Z[f^{=2(N-k_1)}f^{=2(N-k_2)*}]=(N!)^2\delta_{k_1,k_2},
\end{align}
where the average is over the complex random Gaussian matrix.
Here, $(N!)^2$ factor comes by counting the number of orthogonal polynomials in $f^{=2(N-k)}$,
\begin{align}
    \binom{N}{k}^2(k!)^2((N-k)!)^2=(N!)^2,
\end{align}
where $\binom{N}{k}^2$ are from the number of choices for $R,R'$ and $(k!)$ from the coefficients, and $((N-k)!)^2$ from the number of choices for $\sigma',\rho'$.
The noisy probability expression suggests that the high-degree polynomials are more sensitive to the noise and they are suppressed exponentially in their degree.
Also, Eq.~\eqref{eq:ortho} shows that the contribution from high-degree polynomials does not scale as their degree.
Therefore, we will approximate the output probability by truncating the polynomials by setting a cutoff of the degree.
We will show in Sec.~\ref{sec:sample} that the complexity of computing $f^{=2(N-k)}$ is determined by the degree $2(N-k)$.

More concretely, if we choose the maximum degree as $2l$ and truncate higher-degree contributions, we obtain an approximated probability written by the sum of low-degree polynomials
\begin{align}
    \bar{q}(\bm{z})\equiv \sum_{k=N-l}^N x^{N-k} f^{=2(N-k)}.
\end{align}
Then the approximation error is written as
\begin{align}
    \tilde{p}(\bm{z})-\bar{q}(\bm{z})
    =\frac{1}{M^N}\sum_{k=0}^{N-l-1}x^{N-k} f^{=2(N-k)}.
\end{align}

\subsection{Bounds for the total variation distance}
So far, we have focused on the approximation error of a single output probability.
Using this, we will derive the upper bound of the total variation distance of the full probability distribution,
\begin{align}
    \Delta\equiv \sum_{\bm{z}}|\tilde{p}(U,\bm{z})-\bar{q}(U,\bm{z})|
    =\sum_{\bm{z}\in cf}|\tilde{p}(U,\bm{z})-\bar{q}(U,\bm{z})|+|\tilde{p}(U,c)-\bar{q}(U,c)|,
\end{align}
where $cf$ represents the set of all collision-free outcomes and the first sum is over $cf$ and collision outcome $c$.
Here, we explicitly expressed the dependency of $U$.
We note that $\tilde{p}(U,\bm{z})$ is defined as $1-\sum_{\bm{z}\in cf}\tilde{p}(U,\bm{z})$ (see Appendix~\ref{app:kalai}). 
To find the upper bound of the total variation distance, we first need to assign the value of $\bar{q}(U,c)$.
For this moment, let us assign this probability as
\begin{align}
    \bar{q}(U,c)=1-\sum_{\bm{z}\in cf}\bar{q}(U,\bm{z}).
\end{align}
We will show how to make the approximate distribution $\bar{q}$ satisfy the assumption in Appendix~\ref{app:collision}.
Such an assignment makes the analysis much easier because
\begin{align}
    |\tilde{p}(U,c)-\bar{q}(U,c)|
    \leq \sum_{\bm{z}\in cf}|\tilde{p}(U,\bm{z})-\bar{q}(U,\bm{z})|,
\end{align}
which is from the assumption and the triangular inequality.
Then the average squared total variation distance is upper bounded as
\begin{align}
    \mathbb{E}_U[\Delta^2]
    &\leq 4\mathbb{E}_U\left[\left(\sum_{\bm{z}\in cf}|\tilde{p}(U,\bm{z})-\bar{q}(U,\bm{z})|\right)^2\right] \\ 
    &\leq 4\binom{M}{N}\mathbb{E}_U\left[\sum_{\bm{z}\in cf}(\tilde{p}(U,\bm{z})-\bar{q}(U,\bm{z}))^2\right] \\ 
    &\leq 4\binom{M}{N}^2\mathbb{E}_U\left[(\tilde{p}(U,\bm{z})-\bar{q}(U,\bm{z}))^2\right],
\end{align}
where the average is taken over Haar-random unitaries $U$.
We have used Jensen's inequality for the second inequality, and we have used the fact that the average over $U$ gives rise to symmetry to possible collision-free outcomes $\bm{z}\in cf$, the number of which is $\binom{M}{N}$, for the third equality.
By using the low-degree polynomial approximation, its upper bound can be written as
\begin{align}
    \mathbb{E}_U\left[(\tilde{p}(U,\bm{z})-\bar{q}(U,\bm{z}))^2\right]
    &=\frac{1}{M^{2N}}\mathbb{E}_U\left[\left(\sum_{k=0}^{N-l-1}x^{N-k} f^{=2(N-k)}\right)^2\right] \\ 
    &=\frac{(N!)^2}{M^{2N}}\sum_{k=0}^{N-l-1}x^{2(N-k)} \\ 
    &\leq \frac{(N!)^2}{M^{2N}}\sum_{k=0}^{N-l-1}x^{2(l+1)} \\
    &=\frac{(N-l+1)x^{2(l+1)}(N!)^2}{M^{2N}},
\end{align}
where we have used the orthogonality, Eq.~\eqref{eq:ortho}, replacing the Haar-random unitary average with random Gaussian matrix average, for the second equality.
Finally,
\begin{align}
    \mathbb{E}_U\left[\Delta^2\right]
    &\leq 4\binom{M}{N}^2\frac{(N-l+1)x^{2(l+1)}(N!)^2}{M^{2N}} \\ 
    &\leq 4\left(\frac{M^N}{N!}\right)^2\left(N!\right)^{2}\frac{(N-l+1)x^{2(l+1)}}{M^{2N}} \\ 
    &\leq 4Nx^{2(l+1)},
\end{align}
where we have used the inequality $\binom{M}{N}\leq M^N/N!$ for the second inequality.
Together with this, we will use Markov's inequality
\begin{align}
    \text{Pr}_U\left[\Delta \geq \frac{1}{\sqrt{\delta}}\sqrt{\mathbb{E}_U[\Delta^2]}\right]
    =\text{Pr}_U\left[\Delta^2 \geq \frac{1}{\delta}\mathbb{E}_U[\Delta^2]\right]
    \leq \delta,
\end{align}
where the probability is over Haar-random unitary matrices.
Thus for $1-\delta$ portion of Haar-random unitary matrices, the approximation error of low-degree polynomial is upper-bounded by
\begin{align}
    \sum_{\bm{z}}|\tilde{p}(U,\bm{z})-\bar{q}(U,\bm{z})| \leq \frac{2\sqrt{N}x^{l+1}}{\sqrt{\delta}}.
\end{align}
Therefore, to bound the error by $\epsilon>0$, it is sufficient to choose the cutoff of degree $l$ such that
\begin{align}
    l\geq \frac{\log (2\sqrt{N}/\epsilon\sqrt{\delta})}{\log (1/x)}-1
    =O(\log N,\log(1/\epsilon),\log(1/\delta)).
\end{align}
To introduce the noise effect that scales with the system size, we also consider the case where $x$ scales as $x=x_1^\gamma$ with a constant $x_1$.
Then, the total variation distance bound becomes
\begin{align}
    \frac{2\sqrt{N}x^{l+1}}{\sqrt{\delta}}
    =\frac{2\sqrt{N}x_1^{\gamma(l+1)}}{\sqrt{\delta}},
\end{align}
which implies that it is sufficient to choose the degree as
\begin{align}
    l\geq \frac{\log \frac{2\sqrt{N}}{\epsilon\sqrt{\delta}}}{\gamma\log 1/x_1}-1.
\end{align}

\iffalse
Before we present the sampling algorithm, we study the complexity of computing $f^{=2(N-k)}$.
First, the sum over $R$ and $R'$ such that $|R|=|R'|=k$ is taken over $\binom{N}{k}^2$ number of summands.
For fixed $R$ and $R'$, the number of $\sigma'$ and $\tau'$ is $(N-k)!$ for each.
Finally, the product is over $O(N)$ terms.
Therefore, the complexity of computing $f^{=2(N-k)}$ is given by
\begin{align}
    O\left(N\left(\frac{N!}{k!}\right)^2\right).
\end{align}
In other words, the complexity of computing $f^{=2k}$ for a constant $k$ is given by
\begin{align}
    O\left(N\left(\frac{N!}{(N-k)!}\right)^2\right)
    =O(N^{k+1}).
\end{align}
\fi

\iffalse
Let us count the number of possible polynomials we need.
The ingredients are $h_2(z)$ for each $z_{i,j}$ and $z,z^*$; the number of elements for each is $N\times M$.
Suppose that we want to construct $2k$-degree polynomials.
Then the number of possible $2k$-degree polynomials is given by
\begin{align}
    \sum_{j=0}^k \binom{NM}{k}\binom{NM}{k-j}\binom{NM}{k-j}
    \leq \sum_{j=0}^k \binom{NM}{k}\binom{NM}{k}^2
    =O((NM)^{3k}),
\end{align}
where we assumed that $k\leq NM/2$.
\fi

\subsection{Approximate sampling}\label{sec:sample}
In the previous section, we have shown that $\bar{q}(\bm{z})$ with an appropriate cutoff of degree $l$ approximates the noisy distribution $\tilde{p}(\bm{z})$ with an error $\epsilon$ with high probability $1-\delta$.
It is worth emphasizing again that a similar analysis was conducted in ~\cite{kalai2014gaussian} while it focused on approximating a single output probability only and did not provide the bound for total variation distance and a classical approximate sampler of low-degree approximated distribution $\bar{q}(\bm{z})$.
The remaining challenge from the previous section is to find a classical sampling algorithm from $\bar{q}(\bm{z})$.
A caveat is that the approximated distribution $\bar{q}(\bm{z})$ is not necessarily a proper probability distribution, i.e., it might have a negative quantity.
Nevertheless, the following lemma \cite{bremner2017achieving} provides a recipe for dealing with quasi-probability distribution, which can be straightforwardly generalized to $M$-level outcomes instead of binary outcomes:
\begin{lemma} \label{lemma1} (modified)
    Let $\tilde{p}$ be a probability distribution on $M^N$.
    If there is an oracle that computes a function $\bar{q}:M^N\to \mathbb{R}$ as well as its marginals satisfying $\sum_{\bm{x}}\bar{q}(\bm{x})=1$, such that $\|\tilde{p}-\bar{q}\|_1\leq \epsilon$, then there is an algorithm that samples from a probability distribution $q$ using $O(MN)$ calls to the oracle, such that $\|\tilde{p}-q\|_1\leq 2\epsilon$.
\end{lemma}
Here, the marginal is defined as $\bar{q}(x_1,\dots,x_k)=\sum_{x_{k+1},\dots,x_N=1}^M\bar{q}(x_1,\dots,x_N)$.
We have added an additional assumption $\sum_{\bm{x}}\bar{q}(\bm{x})=1$, which results in the approximation error by $2\epsilon$ instead of $4\epsilon/(1-\epsilon)$.
Therefore, it suffices to find $\bar{q}$ whose marginals can be efficiently computed and are close to $\tilde{p}$ so that it can be used for the lemma for noisy Boson Sampling.
The remaining section will show that $\bar{q}$ obtained by sparse low-degree polynomials satisfies such conditions.

One immediate difficulty of applying this lemma to Boson Sampling is that a restriction of an outcome $\bm{z}$ such that $z_1\leq z_2\leq\dots \leq z_N$ makes it difficult to compute its marginals.
To circumvent such a difficulty, we will consider the unordered outcome vector $\bm{r}$ introduced with the first quantization instead of the ordered vector $\bm{z}$.
While the output vector $\bm{r}$ without ordering has a redundancy, it enables us to easily express the marginals since it does not have the restriction of ordering.
Thanks to the symmetry between $\bm{r}$ and $\bm{z}$, we can rewrite it as
\begin{align}
    p(\bm{r})
    =\frac{p(\bm{z})}{N!}
    =\frac{1}{N!}\frac{1}{M^{N}}\sum_{k=0}^N f^{=2(N-k)}(Z),
\end{align}
where $Z$ corresponds to the submatrix of $U$ by choosing the first $N$ rows and $\bm{r}$'s columns.
Our strategy was to set a cutoff on the degree, i.e.,
\begin{align}
    \bar{q}(\bm{r})
    &=\frac{1}{N!}\frac{1}{M^{N}}\sum_{k=N-l}^Nx^{N-k}f^{=2(N-k)}(Z).
\end{align}
Note that changing the representation from $\bm{z}$ to $\bm{r}$ does not change the simulation error due to the symmetry and  Eq.~\eqref{eq:error_relation}.

We now show that marginals can also be computed using a similar method.
The marginal probability of the noiseless distribution is
\begin{align}
    p(r_1,\dots,r_j)
    &=\frac{1}{N!}\sum_{\sigma,\rho\in\mathcal{S}_N} \left(\prod_{i=1}^j U_{\sigma(i),r_i}U^*_{\rho(i),r_i}\right)
    \left(\prod_{i=j+1}^{N} \langle \rho(i)|\sigma(i)\rangle\right)
    \\ 
    &=\frac{1}{N!}\sum_{\substack{J\subset [N]:\\|J|=j}}\sum_{\substack{\tau\in\mathcal{S}_{N-j}: \\ [j+1,N]\to J^c}}\sum_{\substack{\sigma,\rho\in\mathcal{S}_j: \\ [j]\to J}} \left(\prod_{i=1}^j U_{\sigma(i),r_i}U^*_{\rho(i),r_i}\right)
    \left(\prod_{i=j+1}^{N} \langle \tau(i)|\tau(i)\rangle\right) \\ 
    &=\frac{(N-j)!}{N!}\sum_{\substack{J\subset [N]:\\|J|=j}}\sum_{\substack{\sigma,\rho\in\mathcal{S}_j: \\ [j]\to J}} \left(\prod_{i=1}^j U_{\sigma(i),r_i}U^*_{\rho(i),r_i}\right) \\ 
    %&=\frac{1}{M^{j}}\frac{(N-j)!}{N!}\sum_{\substack{J\subset [N]:\\|J|=j}}\sum_{\substack{\sigma,\rho\in\mathcal{S}_j: \\ [j]\to J}} \left(\prod_{i=1}^j z_{\sigma(i),r_i}z^*_{\rho(i),r_i}\right) \\ 
    &=\frac{1}{M^{j}}\frac{(N-j)!}{N!}\sum_{\substack{J\subset [N]:\\|J|=j}}\sum_{\substack{\sigma,\rho\in\mathcal{S}_j:\\ [j]\to J}} \sum_{R\subset T}\left[\prod_{i\in T}h_2(z_{\sigma(i),r_i}) \prod_{i\in T^c}z_{\sigma(i),r_i}z^*_{\rho(i),r_i}\right].
\end{align}
\iffalse
Note that if we count the number of sums, that can be exponentially large, which is from the first sum $\binom{N}{j}$.
However, we have already counted the number of polynomials that are involved in any probabilities, which include marginals since marginals are simply the sum of (possibly exponentially) many probabilities.
It implies that there is redundancy in the sum,
\fi 
Using the same procedure as in the probability case, we can rewrite the noiseless marginal probability as
\begin{align}
    p(r_1,\dots,r_j)
    =\frac{(N-j)!}{N!}\frac{1}{M^{j}}\sum_{k=0}^j g^{=2(j-k)},
\end{align}
where
\begin{align}
    &g^{=2(j-k)} \nonumber
    \\ 
    &=\sum_{\substack{|R|,|R'|=k: \\ R\subset [j],R'\subset[N]}}\sum_{\substack{\sigma\in \mathcal{S}_k:\\
    R\to R'}}\sum_{\substack{K'\subset[N]\setminus R': \\ |K'|=j-k}}\sum_{\substack{\sigma',\rho' \in \mathcal{S}_{j-k}: \\ [j]\setminus R\to K'}}\prod_{i\in S(\sigma',\rho')}h_2(z_{\sigma'(i),r_i})\prod_{i\in ([j]\setminus R)\setminus S(\sigma',\rho')}z_{\sigma'(i),r_i}z^*_{\rho'(i),r_i} \\
    &=\sum_{\substack{|R|,|R'|=k: \\ R\subset [j],R'\subset[N]}}k!\sum_{\substack{K'\subset[N]\setminus R' \\ :|K'|=j-k}}\sum_{\substack{\sigma',\rho' \in \mathcal{S}_{j-k}: \\ [j]\setminus R\to K'}}\prod_{i\in S(\sigma',\rho')}h_2(z_{\sigma'(i),r_i})\prod_{i\in ([j]\setminus R)\setminus S(\sigma',\rho')}z_{\sigma'(i),r_i}z^*_{\rho'(i),r_i} \\
    &=\sum_{|R|=k:R\subset [j]}k!\binom{N}{k}\sum_{\substack{K'\subset[N]: \\ |K'|=j-k}}\sum_{\substack{\sigma',\rho' \in \mathcal{S}_{j-k}: \\ [j]\setminus R\to K'}}\prod_{i\in S(\sigma',\rho')}h_2(z_{\sigma'(i),r_i})\prod_{i\in ([j]\setminus R)\setminus S(\sigma',\rho')}z_{\sigma'(i),r_i}z^*_{\rho'(i),r_i}.
\end{align}
Here $k!$ accounts for the permutations between $R$ and $R'$ and $\binom{N}{k}$ accounts for the choice of $R'$.
Observe that when $j=N$, it reduces to $f^{=2(N-k)}$, which describes the full probability.
Also, the noisy marginal distribution is written as
\begin{align}
    \tilde{p}(r_1,\dots,r_j)
    =\frac{(N-j)!}{N!}\frac{1}{M^{j}}\sum_{k=0}^j x^{j-k} g^{=2(j-k)}.
\end{align}
Thus, the marginal of the approximate distribution is
\begin{align}
    \bar{q}(r_1,\dots,r_j)
    &=\frac{(N-j)!}{N!}\frac{1}{M^{j}}\sum_{k=j-l}^j x^{j-k} g^{=2(j-k)}.
\end{align}
Here, the complexity of computing $g^{=2(j-k)}$ is given by
\begin{align}
    \binom{j}{k}\binom{N}{j-k}((j-k)!)^2
    \leq (Nj)^{j-k}.
\end{align}
Recall that we set a cutoff of the degree as $2(j-k)\leq 2l$.
When $j\leq l$, since the maximum degree is $2j$, we do not approximate and the complexity of computing $\bar{q}(r_1,\dots,r_j)$ is upper-bounded by
\begin{align}
    \sum_{k=0}^j (Nj)^{j-k}
    \leq l(Nl)^{l}\leq N^{2l+1}
    =O(N^{2l+1}),
\end{align}
where we have used $j\leq l\leq N$.
When $j> l$, we start to approximate and the complexity of computing $\bar{q}(r_1,\dots,r_j)$ is given by
\begin{align}
    \sum_{k=j-l}^{j}\binom{j}{k}\binom{N}{j-k}((j-k)!)^2
    \leq (l+1)(Nj)^{l}
    \leq (N+1)(N^2)^{l}
    =O(N^{2l+1}).
\end{align}
Therefore, we can compute any marginals of $\bar{q}(\bm{r})$ in complexity $O(N^{2l+1})$ satisfying ${\|\tilde{p}-\bar{q}\|_1\leq \epsilon}$.
Hence, we can simply apply Lemma~\ref{lemma1} to sample from a proper probability distribution $q$ such that ${\|\tilde{p}-q\|_1\leq 2\epsilon}$.
%Therefore, the complexity of obtaining a sample is $O(N^{2l+2})$ since we need $N$ steps for marginals.

From the previous section, for constant $x$ we showed that the degree $l$ can be chosen to be $l=O\left(\frac{\log (2\sqrt{N}/\epsilon\sqrt{\delta})}{\log (1/x)}\right)$ to bound the total variation distance and that the complexity of computing a single probability (marginal is the same or less) is $O(N^{2l+1})$.
Hence, by using the lemma, the total complexity to generate a sample is then given by
\begin{align}
    N^{O(\log N,\log\epsilon^{-1},\log\delta^{-1})},
\end{align}
which proves Theorem~\ref{thm1}.
As mentioned before, the algorithm's running time is quasi-polynomial not polynomial as in \cite{aharonov2022polynomial}.
The reason is that the noise rate does not scale as the system size for our case.
To properly introduce the noise that scales with the system size, we again consider the case that $x=x_1^\gamma$ with a constant $x_1$.
In this case, $l$ can be chosen to be $l=O\left(\frac{\log\frac{2\sqrt{N}}{\epsilon\sqrt{\delta}}}{\gamma\log 1/x_1}\right)$.
Hence, for $\gamma=\Omega(\log N)$, the complexity becomes polynomial:
\begin{align}
    O(\text{poly}(N,1/\epsilon,1/\delta)),
\end{align}
which proves Corollary~\ref{thm2}.

\iffalse
\begin{align}
    O(N^{2l+1})
    &=O(e^{\frac{\log \frac{4N}{\epsilon^2 \delta}}{2\log 1/x_1}})
    =O\left(\left(\frac{4N}{\epsilon^2 \delta}\right)^{2\log 1/x_1}\right)
    =O(\text{poly}(N,1/\epsilon,1/\delta)).
\end{align}
\fi

We emphasize that the degree of the polynomial of the running time in the noise rate scales as $\log(1/x_1)\approx 1/(1-x_1)$, where the approximation is valid for small noise rate $x_1\approx 1$.
Thus, the running time of our algorithm can be very large due to the large degree of the polynomial, which makes it impractical.
Also, it is worthwhile to emphasize an extreme case where we only choose the lowest degree polynomial, i.e., $l=0$.
Obviously, the lowest degree polynomial, in this case, is a constant, i.e., the corresponding probability distribution is uniform.

\section{Low-degree approximation with partial distinguishability noise}
\subsection{Noise sensitivity and low-degree polynomial approximation}
In various optical experiments including Boson Sampling experiments, one of the most important noise sources is partial distinguishability of particles, which is caused when the particles are not fully indistinguishable because of other degrees of freedom.
The effect of partial distinguishability on Boson Sampling has been studied in~\cite{tichy2015sampling, renema2018efficient, renema2018classical, moylett2019classically}.
Let us study the effect of the noise and approximation method of noisy distribution.

Again, consider an output probability and expand it using an orthogonal polynomial basis
\begin{align}
    p(\bm{z})=|\text{Per}(U_{N,\bm{z}})|^2=\frac{1}{M^{N}}\sum_{\sigma,\rho\in \mathcal{S}_N}\prod_{i=1}^N z_{\sigma(i),i}z^*_{\rho(i),i},
\end{align}
where $Z$ corresponds to a rescaled submatrix of a unitary and is approximated by a random Gaussian matrix.
Then, after introducing the partial distinguishability of photons, the probability becomes \cite{tichy2015sampling}
\begin{align}
    |\text{Per}(Z)|^2=\sum_{\sigma,\rho\in\mathcal{S}_N}\prod_{i=1}^N z_{\sigma(i),i}z^*_{\rho(i),i}
    \to \sum_{\sigma,\rho\in\mathcal{S}_N}x^{N-k}\prod_{i=1}^N z_{\sigma(i),i}z^*_{\rho(i),i},
\end{align}
where $k$ is the number of $i$'s such that $\sigma(i)=\rho(i)$.
In other words, whenever we have an interference due to indistinguishability, i.e., $i\in [M]$ such that $\sigma(i)\neq \rho(i)$, the partial distinguishability $x$ is multiplied as a noise factor ($x=1$ for fully indistinguishable cases and $x=0$ for fully distinguishable cases.).
Now, we expand the probability:
\begin{align}
    \prod_{i=1}^N z_{\sigma(i),i}z^*_{\rho(i),i}
    &=\prod_{i\in T}(z_{\sigma(i),i}z^*_{\sigma(i),i})\prod_{i\in T^c}(z_{\sigma(i),i}z^*_{\rho(i),i})
    =\prod_{i\in T}h_1(z_{\sigma(i),i})\prod_{i\in T^c}h_2(z_{\sigma(i),i},z_{\rho(i),i}).
\end{align}
In this case, we have chosen a different basis of polynomials:
\begin{align}
    1,h_1(z)\equiv zz^*,h_{2}(z,z')\equiv zz'^{*},~~~\text{for independent variables $z$ and $z'$}.
\end{align}
As we have seen, the effect of partial distinguishability is to transform each polynomial as 
\begin{align}
    1\to 1,~~~h_1(z)\to h_1(z),~~~h_2(z)\to xh_2(z).
\end{align}
Here, we assign the degree by adding 0 for $h_1$ and 1 for $h_2$ based on the sensitivity to noise.
Notice a difference from the circuit noise in the previous section that $h_1(z)$ is not sensitive to the noise, and thus it has degree $0$.
We rewrite the summation as
\begin{align}
    |\text{Per}(Z)|^2
    &=\sum_{k=0}^N \sum_{\substack{T,T'\subset [N] \\ |T|=|T'|=k}}\sum_{\substack{\sigma\in \mathcal{S}_k: \\ T\to T'}}\sum_{\substack{\sigma',\rho'\in \mathcal{S}_{N-k}: \\\sigma'(i)\neq \rho'(i), \\ T^c\to T'^{c}}}\prod_{i\in T}h_1(z_{\sigma(i),i})\prod_{i\in T^c}h_2(z_{\sigma'(i),i},z_{\rho'(i),i})
    =\sum_{k=0}^N f^{=(N-k)},
\end{align}
where $k$ is the number of $i$'s such that $\sigma(i)=\rho(i)$ from the previous notation.
For each $k$, we need to decide $k$ elements from $[N]$ for input and output, which are represented by $T$ and $T'$.
The new $\sigma$ is the permutation between these newly chosen sets.
And $\sigma'$ and $\rho'$ are now permutations between the remaining $(N-k)$ indices and $\sigma'(i)\neq \rho'(i)$ for all $i$'s.
Thus, the $(N-k)$th degree part is written as
\begin{align}
    f^{=(N-k)}
    &=\sum_{\substack{T,T'\subset [N] \\ |T|=|T'|=k}}\sum_{\substack{\sigma\in \mathcal{S}_k: \\ T\to T'}}\sum_{\substack{\sigma',\rho'\in \mathcal{S}_{N-k}: \\\sigma'(i)\neq \rho'(i), \\ T^c\to T'^{c}}}\prod_{i\in T}h_1(z_{\sigma(i),i})\prod_{i\in T^c}h_2(z_{\sigma'(i),i},z_{\rho'(i),i}).
\end{align}

After some algebra, we can show that (See Appendix~\ref{app:dist})
\begin{align}
    \mathbb{E}_Z[f^{=(N-k_1)}f^{=(N-k_2)*}]
    =0, ~~~ \text{if}~~k_1\neq k_2,
\end{align}
and that
\begin{align}
    \mathbb{E}_Z[|f^{=(N-k)}|^2]
    =\binom{N}{k}^2(N-k)!(!(N-k))\sum_{j=0}^k\binom{k}{j}^2j!(k-j)!(!(k-j))2^j,
\end{align}
where $(!k)$ represents the number of derangements of $k$ elements, i.e., the number of permutations $\sigma$ between $k$ elements such that $\sigma(i)\neq i$ for any $i\in[k]$.
When the photons in the system have partial distinguishability $x$, the polynomial transforms as
\begin{align}
    f^{=(N-k)}\to x^{N-k}f^{=(N-k)}.
\end{align}
Thus, our approximation strategy is to keep the polynomials up to degree $l$:
\begin{align}
    \tilde{p}(\bm{z})
    =\frac{1}{M^{N}}\sum_{k=0}^N x^{N-k}f^{=(N-k)} 
    \approx \frac{1}{M^{N}}\sum_{k=N-l}^N x^{N-k}f^{=(N-k)} 
    \equiv \bar{q}(\bm{z}),
\end{align}
and the approximation error is
\begin{align}
    \tilde{p}(\bm{z})-\bar{q}(\bm{z})
    =\frac{1}{M^N}\sum_{k=0}^{N-l-1}x^{N-k}f^{=(N-k)}.
\end{align}
%What is the complexity of computing the approximation?
%For fixed degree $l$, the number of $T$ and $T'$ is $\binom{N}{N-l}^2$, which is polynomial in $N$.
%And the number of the permutations $\sigma,\sigma',\rho'$ for a constant $l$ is only polynomial in $N$.

\subsection{Bounds for the total variation distance}
Using the same method as the previous section, we can show that
\begin{align}
    \mathbb{E}_U[\Delta^2]\leq 4\binom{M}{N}^2\mathbb{E}_U\left[\tilde{p}(U,\bm{z})-\bar{q}(U,\bm{z})\right]^2
    =4\binom{M}{N}^2\frac{1}{M^{2N}}\sum_{k=l+1}^{N}x^{2k}\mathbb{E}_Z[|f^{=k}|^2].
\end{align}
In Appendix~\ref{app:dist}, we show that
\begin{align}
    \mathbb{E}_Z[|f^{=k}|^2]\leq e^2 (N!)^2.
\end{align}
(Note that one can numerically check that $e^2$ is generally not necessary \cite{renema2018efficient} but we keep it since it does not change our main result below.)
Hence, the average squared total variation distance is bounded as
\begin{align}
    \mathbb{E}_U[\Delta^2]
    \leq 4\binom{M}{N}^2\frac{1}{M^{2N}}\sum_{k=l+1}^{N}x^{2k} e^2(N!)^2
    \leq 4\sum_{k=l+1}^{N}x^{2(l+1)} e^2
    \leq 4e^2Nx^{2(l+1)}.
\end{align}
By applying Markov's inequality as the previous case, we can conclude that for $1-\delta$ portion of Haar-random linear-optical circuits, the approximation error of low-degree polynomial is upper-bounded by
\begin{align}
    \sum_{\bm{z}}|\tilde{p}(U,\bm{z})-\bar{q}(U,\bm{z})|\leq \frac{2e \sqrt{N}x^{l+1}}{\sqrt{\delta}}.
\end{align}
To bound the error by $\epsilon$, it is sufficient to choosse $l$ to be
\begin{align}
    l= \frac{\log\left(\frac{2e\sqrt{N}}{\epsilon\sqrt{\delta}}\right)}{\log(1/x)}-1
    =O(\log N,\log(1/\epsilon),\log(1/\delta)).
\end{align}

\subsection{Barrier of approximate sampling}
Now, we again try to find an analogous classical sampler to the previous case and show a barrier to implementing it in an efficient way.
First of all, the noisy distribution is written as
\begin{align}
    \tilde{p}(\bm{r})=\frac{1}{M^NN!}\sum_{k=0}^N x^{N-k}f^{=(N-k)}.
\end{align}
Our strategy was to set a cutoff $l$ on the degree, i.e.,
\begin{align}
    \bar{q}(\bm{r})=\frac{1}{M^NN!}\sum_{k=N-l}^N x^{N-k}f^{=(N-k)}.
\end{align}
One can easily check that the number of summands in $f^{(N-k)}$ is given by
\begin{align}
    \binom{N}{k}^2 k! (N-k)! (!(N-k)),
\end{align}
which is larger than $N!$ regardless of $k$.
Thus, direct computation of $f^{=(N-k)}$ is inefficient to any degrees.
One might hope that there can still be a possibility of computing this quantity efficiently.
However, we can show that exact computation requires exponential time.
To see this, consider the lowest-degree polynomial $l=0$, which is the fixed point of the noise:
\begin{align}
    \sum_{\sigma \in\mathcal{S}_N} \left(\prod_{i=1}^N U_{\sigma(i),r_i}U^*_{\sigma(i),r_i}\right)
    =\per (|U_{N,\bm{r}}|^2),
\end{align}
where $|U|^2$ is the matrix obtained by taking absolute values on each matrix element.
Therefore, it is written as the permanent of a positive matrix, and its exact computation is known to be \#P-hard \cite{valiant1979complexity}.
Meanwhile, \cite{renema2018efficient} observed that the permanent of positive matrices can be efficiently approximated in multiplicative error \cite{jerrum2004polynomial}.
Let us recall their method and present a caveat.
We can rewrite the polynomial as in \cite{renema2018efficient}:
\begin{align}
    f^{=(N-k)}
    &=\sum_{\substack{T,T'\subset [N] \\ |T|=|T'|=k}}\sum_{\substack{\sigma\in \mathcal{S}_k: \\ T\to T'}}\sum_{\substack{\sigma',\rho'\in \mathcal{S}_{N-k}: \\\sigma'(i)\neq \rho'(i), \\ T^c\to T'^{c}}}\prod_{i\in T}h_1(z_{\sigma(i),i})\prod_{i\in T^c}h_2(z_{\sigma'(i),i},z_{\rho'(i),i}) \\ 
    &=\sum_{\substack{T,T'\subset [N] \\ |T|=|T'|=k}}\per (|Z_{T',T}|^2)\sum_{\substack{\tau'\in \mathcal{S}_{N-k}: \\  \tau'(i)\neq i\\ T'^c\to T'^{c}}}\sum_{\substack{\sigma'\in \mathcal{S}_{N-k}: \\  T^c\to T'^{c}}}\prod_{i\in T^c}h_2(z_{\sigma'(i),i},z_{\tau'(\sigma'(i)),i}) \\
    &=\sum_{\substack{T,T'\subset [N] \\ |T|=|T'|=k}}\per (|Z_{T',T}|^2)\sum_{\substack{\tau'\in \mathcal{S}_{N-k}: \\  \tau'(i)\neq i\\ T'^c\to T'^{c}}}\sum_{\substack{\sigma'\in \mathcal{S}_{N-k}: \\  T'^c\to T^{c}}}\prod_{i\in T'^c}h_2(z_{i,\sigma'(i)},z_{\tau'(i),\sigma'(i)}) \\ 
    &=\sum_{\substack{T,T'\subset [N] \\ |T|=|T'|=k}}\sum_{\substack{\tau'\in \mathcal{S}_{N-k}: \\  \tau'(i)\neq i\\ T^c\to T'^{c}}}\per (|Z_{T',T}|^2)\per(Z_{T'^c,T^c}*Z_{\tau'(T'^c),T^c}),
\end{align}
where $*$ represents the elementwise multiplication of two matrices and $Z_{T'^c,T^c}$ is obtained by selecting rows and columns corresponding to $T'^c$ and $T^c$, respectively, and $Z_{\tau'(T'^c),T^c}$ is obtained similarly but with permuting the rows by $\tau'$.
One can notice that if we set $N-k=l$, the number of terms to sum is
\begin{align}
    \binom{N}{N-l}^2 (!l),
\end{align}
which is a polynomial in $l$.
Also, the matrix size of $Z_{T'^c,T^c}*Z_{\tau'(T'^c),T^c}$ is given by $l\times l$, whose permanent can be exactly computed in $\tilde{O}(2^l)$ \cite{ryser1963combinatorial}.
Meanwhile, the difficulty comes from computing the permanent of $|Z_{T',T}|^2$, whose matrix size is $(N-l)\times (N-l)$.
\cite{renema2018efficient} claimed that since we can efficiently approximate the permanent of positive matrices in multiplicative error \cite{jerrum2004polynomial}, it might enable us to approximate $f^{=l}$ as well.
However, this is not immediately obvious.
To see this clearer, we can simply write $f^{=l}$ as an inner product of two vectors $\bm{a},\bm{b}\in \mathbb{C}^{\poly(N)}$,
\begin{align}
    f^{=l}=\bm{a}\cdot \bm{b},
\end{align}
where all the elements of $\bm{a}$ can be exactly computed and those of $\bm{b}$ can be efficiently approximated in multiplicative error, which corresponds to $\per(Z_{T'^c,T^c}*Z_{\tau'(T'^c),T^c})$.
The difficulty is the fact that even though we have exact values of $\bm{a}$, they can be negative (or even complex).
Thus, the quantity $f^{=l}$ we are approximating is the sum of many terms, which can be only be approximated, with different signs.
In general, it does not guarantee even a multiplicative error approximation for $f^{=l}$.
Therefore, the difficulty of computing the probability becomes a barrier to applying the same technique to partial distinguishability even though the approximation error using low-degree polynomials is sufficiently small.
Furthermore, even if we could approximate the probabilities in a multiplicative error, the direct application of Lemma~\ref{lemma1} still requires an exact computation of probabilities and marginals.

Our analysis reveals that channeling between the small approximation error (in total variation distance) and constructing an efficient classical sampler is highly nontrivial.
More precisely, our analysis implies that an additional condition is required for noise, which is that the low-degree polynomials need to be composed only of polynomially many orthogonal basis polynomials.

We remark that even though the probability of the fixed point of partial distinguishability noise, i.e., fully distinguishable Boson Sampling, is described by the permanent of a positive matrix and computing the probability is hard, the corresponding sampling can be shown to be easy even exactly \cite{aaronson2011computational, aaronson2013bosonsampling}.
This is because fully distinguishable particles do not interfere, so that we can sample particle by particle, which does not require computing the probability of $N$ particles.
Therefore, it remains open to adapt such a method without computing probabilities to circumvent the barrier and construct an approximate sampler.

%Again, it is worth emphasizing that a similar analysis for approximation error was conducted in ~\cite{renema2018efficient} while it did not provide a provable classical approximate sampler of low-degree approximated distribution $\bar{q}(\bm{r})$.

\iffalse
The marginal probability of the noisy distribution is
\begin{align}
    \tilde{p}(r_1,\dots,r_k)
    &=\frac{1}{N!}\sum_{\sigma,\rho\in\mathcal{S}_N} x^{\omega(\sigma,\rho)}\left(\prod_{i=1}^k U_{\sigma(i),r_i}U^*_{\rho(i),r_i}\right)
    \left(\prod_{i=k+1}^{N} \langle \rho(i)|\sigma(i)\rangle\right).
\end{align}
And the marginal distribution of the approximate distribution is
\begin{align}
    \bar{q}(r_1,\dots,r_k)
    &=\frac{1}{N!}\sum_{\sigma,\rho\in\mathcal{S}_N:\omega(\sigma,\rho)\leq l} x^{\omega(\sigma,\rho)}\left(\prod_{i=1}^k U_{\sigma(i),r_i}U^*_{\rho(i),r_i}\right)
    \left(\prod_{i=k+1}^{N} \langle \rho(i)|\sigma(i)\rangle\right).
\end{align}
\fi

\iffalse
Finally, let us again consider an extreme case where we only choose the lowest degree polynomial, i.e., $l=0$.
Remarkably, in this case, the lowest degree polynomial is not a constant.
More precisely, the polynomial corresponds to choosing $\sigma=\rho$, which is the probability of distinguishable Boson Sampling.
\fi

\section{Barriers to photon Loss}
Finally, let us consider photon-loss which is one of the most detrimental noise models in Boson Sampling experiments.
We can assume that all the loss occurs at the beginning with total transmission rate $\eta=\eta_1^{d}$, where $d$ is the depth of the circuit and $\eta_1$ is a constant loss rate per depth.
This simplification can be justified in many cases because uniform loss channel and beam splitters commute.

In the second quantization representation, the density matrix of the state is written as
\begin{align}
    |1,\dots,1,0\dots,0\rangle\langle 1,\dots,1,0\dots,0|,
\end{align}
which represents the number of photons for each mode.
The effect of photon loss is to transform a single-photon state as
\begin{align}
    |1\rangle\langle 1|\to \eta|1\rangle\langle 1|+(1-\eta)|0\rangle\langle0|
\end{align}
and the vacuum state $|0\rangle\langle 0|$ does not change.
Therefore, if we introduce photon loss, the state transforms
\begin{align}
    \sum_{k=0}^N \binom{N}{k}\eta^k(1-\eta)^{N-k}\hat{\rho}_k,
\end{align}
where $\hat{\rho}_k$ is $k$-photon states with equal weight of selecting $k$ photons out of the initial $N$ photons.
One distinct feature of photon loss from other noise models is that the photon number changes and that the output quantum state occupies lower than $N$ photons.

If we exploit the same method as the previous cases, we will need to discard the terms having $\eta^k$ with $k>l$ with a cutoff $l$.
It implies that we discard
\begin{align}
    \sum_{k=l+1}^N \binom{N}{k}\eta^k(1-\eta)^{N-k}\hat{\rho}_k,
\end{align}
which contains at least $\eta^{l+1}$ degrees, while there are other remaining terms that contain $\eta^{l+1}$ degrees; thus, we will underestimate the approximation error.
We note that by discarding the above term, we do not obtain any outcomes which have larger than $l$ photons because Boson Sampling circuit does not change the number of photons.
Even when underestimating the approximation error, one can easily see that the probability of the discarded terms is given by
\begin{align}
    \text{Tr}\left[\sum_{k=l+1}^N \binom{N}{k}\eta^k(1-\eta)^{1-k}\hat{\rho}_k\right]
    =\sum_{k=l+1}^N \binom{N}{k}\eta^k(1-\eta)^{N-k}.
\end{align}
Here, we emphasize that $\hat{\rho}_k$'s for different $k$'s are orthogonal each other from the density matrix level, which is a distinct property from the other noise models.
Thus, regardless of a linear-optical circuit, the probability that we have lost from discarding high-degree contributions of $\eta$ is already large.
To be more precise, notice that the photon number distribution follows the binomial distribution with mean $\eta N$ and standard deviation $\sqrt{N\eta(1-\eta)}$.
It suggests that we need to keep at least $l\geq \eta N$.
As a comparison, for circuit noise and partial distinguishability, the required degree was $l=O(\log N)$ for a constant noise rate, which shows that the required degree for photon loss is much larger.

Now, let us now consider the output probability of obtaining $\bm{r}$ which has $k$ clicks with $N-k$ photons lost and analyze the complexity.
Without loss of generality, let us set $r_i=0$ for $k+1\leq i \leq N$.
Then, the output probability of lossy Boson Sampling is written as
\begin{align}
    \tilde{p}(\bm{r})
    &=\frac{\eta^{k}(1-\eta)^{N-k}}{N!}\binom{N}{k}^{-1}\sum_{T\subset[N]:|T|=k}|\per(U_{T,\bm{r}})|^2.
    %&=\frac{\eta^{k}(1-\eta)^{N-k}}{N!}\binom{N}{k}^{-1}\sum_{T\subset[N]:|T|=k}\sum_{\substack{\sigma,\rho\in \mathcal{S}_k: \\ [k]\to T}}\prod_{i=1}^k U_{\sigma(i),r_i}U_{\rho(i),r_i}^* \\ 
    %&=\frac{\eta^{k}(1-\eta)^{N-k}}{N!M^{N-k}}\binom{N}{k}^{-1}\sum_{T\subset[N]:|T|=k}\sum_{\substack{\sigma,\rho\in \mathcal{S}_k: \\ [k]\to T}}\prod_{i=1}^k z_{\sigma(i),r_i}z_{\rho(i),r_i}^*.
\end{align}
Approximating by low-degree in $\eta$ only changes the prefactor as
\begin{align}\label{eq:sum_per}
    \bar{q}(\bm{r})
    &=
    \frac{\eta^{k}}{N!}\binom{N}{k}^{-1}\sum_{j=0}^{l-k}\binom{N-k}{j}(-\eta)^j\sum_{T\subset[N]:|T|=k}|\per(U_{T,\bm{r}})|^2.
\end{align}
Thus, the complexity of $\bar{q}(\bm{r})$ by computing all the permanents and summing them is
\begin{align}
    \tilde{O}\left(\binom{N}{k}2^{k}\right)
    =\tilde{O}\left(N^{k}\right),
\end{align}
which is exponential in $k$.
Therefore, to make the complexity at most quasi-polynomial as before, $l$ needs to be at most logarithmic in the system size $N$, $l=O(\log N)$, which requires the condition $\eta N=O(\log N)$.

However, it is known that when $\eta N=O(\sqrt{N})$, the corresponding noisy distribution can be approximated by a separable state or thermal state input Boson Sampling \cite{oszmaniec2018classical, garcia2019simulating}, which can be easily simulated using a classical computer.
More specifically, the trace distance between lossy single photons and a thermal state converges to 0 when $\eta N=o(\sqrt{N})$ in an asymptotic regime (it converges to a constant when $\eta N=\Theta(\sqrt{N})$).
Therefore, the regime in which the proposed technique might work can already be classically simulated using different techniques with the approximation error converging to zero in the asymptotic regime.

It is worth emphasizing that we assumed that the sum of permanents Eq.~\eqref{eq:sum_per} can only be obtained by computing individual permanents, which might not be the optimal method.
For certain cases, exponential sum of quantities that are hard to compute can be easily obtained \cite{oh2022quantum}.

\iffalse
\begin{figure*}[t]
\includegraphics[width=400px]{scheme_mod.eps} %{scheme_ed.eps}
\caption{Molecular vibronic spectra generation based on Gaussian Boson Sampling and the proposed classical algorithm using the positive $P$-representation to obtain Fourier components of the spectra. 
Our main result is to find an analytic solution of Fourier components of molecular vibronic spectra, which enables us to generate the spectra by using the inverse Fourier transformation.}
\label{fig:scheme}
\end{figure*}
\fi

\section*{Acknowledgements}
We thank Senrui Chen and Umesh Vazirani for interesting and fruitful discussions.  
LJ acknowledges support from the ARO MURI (W911NF-21-1-0325), AFOSR MURI (FA9550-19-1-0399, FA9550-21-1-0209), AFRL (FA8649-21-P-0781), DoE Q-NEXT, NSF (OMA-1936118, ERC-1941583, OMA-2137642), NTT Research, and the Packard Foundation (2020-71479).
BF  acknowledges support from AFOSR (YIP number FA9550-18-1-0148 and
FA9550-21-1-0008). This material is based upon work partially
supported by the National Science Foundation under Grant CCF-2044923
(CAREER) and by the U.S. Department of Energy, Office of Science,
National Quantum Information Science Research Centers as well as by
DOE QuantISED grant DE-SC0020360.

\bibliographystyle{alpha}
\bibliography{reference.bib,fefferman.bib}

\appendix

\section{Output distribution of Gaussian noise}\label{app:kalai}
In this Appendix, we show that the output probability distribution after Gaussian noise \cite{kalai2014gaussian} is a nontrivial and proper probability distribution.
Specifically, we show that the probability of obtaining outcomes in the collision-free subspace is close to and smaller than one, as in the noiseless case.
Therefore, by defining the remaining probability to normalize the sum of probabilities, the noise maps a noiseless output probability distribution into another proper probability distribution.
Recall that Gaussian noise transforms the unitary matrix of a boson sampling circuit as 
\begin{align}
    U \to \sqrt{x}U+\sqrt{1-x}Y,
\end{align}
where $Y$ is a random Gaussian matrix with variance $1/M$.
Consider a probability of detecting $N$ photons for the first $N$ modes with $N$ input photons from the first $N$ modes:
\begin{align}
    p(U,\bm{z})=|\per(U_{N,N})|^2
    =\sum_{\sigma,\rho\in\mathcal{S}_N}\prod_{i=1}^N U_{i,\rho(i)}U_{i,\sigma(i)}^*.
\end{align}

After Gaussian noise, it transforms to
\begin{align}
    \tilde{p}(U,\bm{z})
    &=\mathbb{E}_Y\left[\sum_{\sigma,\rho\in\mathcal{S}_N}\prod_{i=1}^N (\sqrt{x}U_{i,\rho(i)}+\sqrt{1-x}Y_{i,\rho(i)})(\sqrt{x}U_{i,\sigma(i)}+\sqrt{1-x}Y_{i,\sigma(i)})^*\right] \\ 
    &=\mathbb{E}_Y\left[\sum_{\sigma,\rho\in\mathcal{S}_N}\prod_{i=1}^N (x U_{i,\sigma(i)}U_{i,\rho(i)}^*+(1-x)Y_{i,\sigma(i)}Y_{i,\rho(i)}^*)\right] \\ 
    &=\sum_{k=0}^N \frac{x^k(1-x)^{N-k}}{M^{N-k}}(N-k)!\sum_{\substack{K,K'\subset [N]: \\ |K|=|K'|=k}}\sum_{\sigma,\rho\in\mathcal{S}_k:K \to K'}\prod_{i\in K}  U_{i,\sigma(i)}U_{i,\rho(i)}^* \\ 
    &=\sum_{k=0}^N \frac{x^k(1-x)^{N-k}}{M^{N-k}}(N-k)!\sum_{\substack{K,K'\subset [N]: \\ |K|=|K'|=k}}|\per (U_{K,K'})|^2,
\end{align}
where for the second equality, we used the independence of matrix elements of $Y$, and for the third equality, we split permutations into trivial permutations, from again independence of $Y$'s elements, and nontrivial permutations.
Let us sum over all collision-free outcomes $\bm{z}$:
\begin{align}
    \sum_{\bm{z}\in cf}\tilde{p}(U,\bm{z}) 
    &=\sum_{\bm{z}\in cf}\sum_{k=0}^N \frac{x^k(1-x)^{N-k}}{M^{N-k}}(N-k)!\sum_{\substack{K\subset [N]: \\ |K|=k}}\sum_{\substack{K'\subset \bm{z}: \\ |K'|=k}}|\per (U_{K,K'})|^2 \\ 
    &=\sum_{k=0}^N \frac{x^k(1-x)^{N-k}}{M^{N-k}}\binom{M-k}{N-k}(N-k)!\sum_{\substack{K\subset [N]: \\ |K|=k}}\sum_{\substack{K'\subset [M]: \\ |K'|=k}}|\per (U_{K,K'})|^2 \\ 
    &=\sum_{k=0}^N \frac{x^k(1-x)^{N-k}}{M^{N-k}}\frac{(M-k)!}{(M-N)!}\sum_{\substack{K\subset [N]: \\ |K|=k}}\text{(collision-free for $K$ input boson sampling with $U$)} \\ 
    &=\sum_{k=0}^N \frac{x^k(1-x)^{N-k}}{M^{N-k}}\frac{(M-k)!}{(M-N)!}\sum_{\substack{K\subset [N]: \\ |K|=k}}[1-\text{(collision for $K$ input boson sampling with $U$)}],
\end{align}
where (collision(-free) for $K$ input boson sampling with $U$) represents the probability of obtaining collision(-free) outcomes with $|K|$ single photons in modes $K$ with the circuit unitary $U$, and the inclusion symbol from $\bm{z}$ is defined to be the subsets of the modes $i$'s such that $z_i=1$.
First, we find the upper bound:
\begin{align}
    &\sum_{k=0}^N \frac{x^k(1-x)^{N-k}}{M^{N-k}}\frac{(M-k)!}{(M-N)!}\sum_{\substack{K\subset [N]: \\ |K|=k}}[1-\text{(collision for $K$ input boson sampling with $U$)}] \\ 
    &<
    \sum_{k=0}^N x^k(1-x)^{N-k}\binom{N}{k}
    =1.
\end{align}
Now we find the lower bound of the average over Haar-random unitary $U$.
Using the bosonic birthday paradox \cite{aaronson2011computational}, we can bound the total collision-free outcomes as
\begin{align}
    &\mathbb{E}_U\left[\sum_{\bm{z}\in cf}\tilde{p}(U,z)\right] \\ 
    &=
    \mathbb{E}_U\left[\sum_{k=0}^N \frac{x^k(1-x)^{N-k}}{M^{N-k}}\frac{(M-k)!}{(M-N)!}\sum_{\substack{K\subset [N]: \\ |K|=k}}[1-\text{(collision for $K$ input boson sampling with $U$)}]\right] \\ 
    &> \sum_{k=0}^N \frac{x^k(1-x)^{N-k}}{M^{N-k}}\frac{(M-k)!}{(M-N)!}\sum_{\substack{K\subset [N]: \\ |K|=k}}\left(1-\frac{2k^2}{M}\right) \\ 
    &\geq \sum_{k=0}^Nx^k(1-x)^{N-k}\binom{N}{k}\left(1-\frac{N}{M}\right)^{N}\left(1-\frac{2N^2}{M}\right) \\ 
    &\to 1,
\end{align}
where for the last expression, we used $M=\omega(N^2)$ for large $N$.
Therefore, using the assumption of the strong collision-free regime, i.e., $M=\omega(N^5)$, the noisy output probability distribution sums close to one.
Finally, we defined the collision case of the noisy distribution as the remaining probability, so that the total probability is normalized to be one,
\begin{align}
    \sum_{\bm{z}\in cf}\tilde{p}(U,\bm{z})+\tilde{p}(U,c)=1.
\end{align}

\section{Collision}\label{app:collision}
\iffalse
To make this, consider a marginal
\begin{align}
    \bar{q}(r_1,\dots,r_j)
    &=\frac{(N-j)!}{N!}\frac{1}{M^{j}}\sum_{k=j-l}^j x^{j-k} g^{=2(j-k)}.
\end{align}
Suppose $\{r_i\}_{i=1}^{j-1}$ are distinct, i.e. collision-free.
And now we need to consider $r_j$ which can generate collision outcome.
Then, for $r_j$ corresponding to a collision-free outcome, i.e., $r_j$ is different from $\{r_i\}_{i=1}^{j-1}$, we will define
\begin{align}
    \bar{q}(r_1,\dots,r_{j-1},c)
    \equiv \bar{q}(r_1,\dots,r_{j-1})-\sum_{r_j\in [M]\setminus \{r_i\}_{i=1}^{j-1}}\bar{q}(r_1,\dots,r_{j}).
\end{align}
We do this for every step.
Then, it guarantees that the condition
\begin{align}
    \bar{q}(c)+\sum_{\bm{r}\in c}\bar{q}(\bm{r})=1.
\end{align}

Also, it is already guaranteed that
\begin{align}
    \sum_{r_1=1}^M\bar{q}(r_1)=1.
\end{align}
\fi

%Note that thanks to bosonic birthday paradox~\cite{aaronson2011computational}, the possible error arising from discarding collision outcomes is suppressed by inverse polynomially $\epsilon_c=O(N^2/M)$ which converges to 0 in asymptotic regime.

In this Appendix, we will show how to make the distribution $\bar{q}(\bm{r})$ to satisfy the sufficient condition $\sum_{\bm{r}\in[M]^N}\bar{q}(\bm{r})=1$ by assigning $\bar{q}(\bm{r})$ for collision cases $\bm{r}$ properly.
We will assume that we have chosen the cutoff of degree as $l\geq 1$ for simplicity.
Then, the first-order marginal $\bar{q}(r_1)$ is exact, i.e., $\bar{q}(r_1)=\tilde{p}(r_1)$ for all $r_1\in[M]$.
For the second-order marginals,
we will define for each $r_1\in [M]$
\begin{align}
    \bar{q}(r_1,r_2=r_1)=\bar{q}(r_1)\left[1-\sum_{r_2\in [M]\setminus \{r_1\}}\bar{q}(r_1,r_2)\right],
\end{align}
which obviously guarantees that $\sum_{r_2=1}^M \bar{q}(r_1,r_2)=\bar{q}(r_1)$.
Similarly, for given $(r_1,\dots,r_{k-1})$ with distinct $\{r_i\}_{i=1}^{k-1}$, we define for each $r_k \in \{r_i\}_{i=1}^{k-1}$
\begin{align}
    \bar{q}(r_1,\dots,r_k)
    =\bar{q}(r_1,\dots,r_{k-1})\left[1-\frac{1}{k-1}\sum_{r_k\in [M]\setminus \{r_i\}_{i=1}^{k-1}} \bar{q}(r_1,\dots,r_{k-1},r_k)\right]~\text{for each}~ r_k \in \{r_i\}_{i=1}^{k-1},
\end{align}
which again guarantees that $\sum_{r_k=1}^M \bar{q}(r_1,\dots,r_k)=\bar{q}(r_1,\dots,r_{k-1})$.
For such $r_k$'s and for all permutations $\sigma\in\mathcal{S}_k$, we also define
\begin{align}
    \bar{q}(r_{\sigma(1)},\dots,r_{\sigma(k)})
    \equiv \bar{q}(r_1,\dots,r_k).
\end{align}
We continue this procedure until $k=N$ when we define all quantities of $\bar{q}(\bm{r})$ of $\bm{r}\in[M]^N$.

Now, we have defined all relevant quantities of $\bar{q}(\bm{r})$ and its marginals.
Consequently, we can easily show that the resultant distribution satisfies
\begin{align}
    \sum_{\bm{r}\in [M]^N}\bar{q}(\bm{r})=1,
\end{align}
which can be easily shown by the marginal relation,
\begin{align}
    \bar{q}(r_1,\dots,r_{k-1})
    =\sum_{r_k=1}^M\bar{q}(r_1,\dots,r_{k}).
\end{align}

As a remark, we argue why this procedure is necessary.
Since the collision probability is inverse-polynomially suppressed when $M=\omega(N^2)$ \cite{aaronson2011computational}, one might be tempted to set it to be zero for $\bar{q}(\bm{r})$.
However, one can immediately see that it might cause a large error.
Suppose that a quasi-probability distribution $\bar{q}(\bm{r})$ is given for collision-free space, i.e., which is close to the target distribution
\begin{align}
    \sum_{\bm{r}\in cf}|\tilde{p}(\bm{r})-\bar{q}(\bm{r})|\leq \epsilon,
\end{align}
where $cf$ accounts for the set of collision-free outcomes.
We first show that a naive approach may entail a large error.
Let us denote the probability of collisions as $\epsilon_c$.
We will set $\bar{q}(\bm{r})=0$ for collision outcomes $\bm{r}$.
Then, for full distribution we have
\begin{align}
    \sum_{\bm{r}\in [M]^N}|\tilde{p}(\bm{r})-\bar{q}(\bm{r})|
    =\sum_{\bm{r}\in c}|\tilde{p}(\bm{r})-\bar{q}(\bm{r})|
    +\sum_{\bm{r}\in cf}|\tilde{p}(\bm{r})-\bar{q}(\bm{r})|
    \leq \epsilon_c+\epsilon\equiv \epsilon_t,
\end{align}
where $c$ accounts for the set of collision outcomes.
Then after using the lemma from \cite{bremner2017achieving}, we can sample from a proper probability distribution $q(\bm{z})$ with the total variation distance given by
\begin{align}
    \sum_{\bm{z}}|\tilde{p}(\bm{z})-q(\bm{z})|
    \leq \frac{4\epsilon_t}{1-\epsilon_t}.
\end{align}
Since the collision probability $\epsilon_c$ is fixed for a given system, we cannot reduce the error as much as we want.
Thus, we need to assign appropriate quantities of $\bar{q}(\bm{r})$ for collision outcomes before applying the lemma.

\section{Orthogonality of polynomials for partial distinguishability noise}\label{app:dist}
In this Appendix, we show the orthogonality of polynomials introduced for partial distinguishability noise.
Consider
\begin{align}
    \mathbb{E}_Z[|f^{=(N-k)}|^2]
    &=\mathbb{E}_Z\bigg[\left(\sum_{T,T'\subset [N],|T|=|T'|=k}\sum_{\sigma,\sigma',\rho'}\prod_{i\in T}h_1(z_{\sigma(i),i})\prod_{i\in T^c}h_2(z_{\sigma'(i),i},z_{\rho'(i),i})\right) \nonumber \\ 
    &~~~~~\times \left(\sum_{T^*,T'^{*}\subset [N],|T^*|=|T'^*|=k}\sum_{\sigma^*,\sigma'^*,\rho'^*}\prod_{i\in T^*}h_1^*(z_{\sigma^*(i),i})\prod_{i\in T^{*c}}h_2^*(z_{\sigma'^*(i),i},z_{\rho'^*(i),i})\right)\bigg].
\end{align}
Here if $T\neq T^*$ or $T'\neq T^{'*}$, one can easily check that the average over $Z$ becomes zero.
Thus, we set $T^*=T$ and $T'^{*}=T'$.
Now, we have, for $|T|=k$ and a fixed $T$ and $T'$,
\begin{align}
    \sum_{\sigma,\sigma^*}\prod_{i\in T}h_1(z_{\sigma(i),i})h_1^*(z_{\sigma^*(i),i})
    =\sum_{j=0}^k\binom{k}{j}j!(k-j)!(!(k-j))2^j,
\end{align}
which can be shown by splitting the factors $|z|^4$ and $|z|^2$ and counting each and using $\mathbb{E}_z[|z|^4]=2$.
Here $(!j)$ is the derangement, namely, the number of permutations $\sigma\in\mathcal{S}_j$ such that $\sigma(i)\neq i$ for all $i$'s.
Meanwhile,
\begin{align}
    \sum_{\sigma',\rho'}\sum_{\sigma'^{*},\rho'^{*}}
    \prod_{i\in T^c}h_2(z_{\sigma'(i),i},z_{\rho'(i),i})h_2^*(z_{\sigma^{'*}(i),i},z_{\rho^{'*}(i),i})
    &=\sum_{\sigma',\rho'}\sum_{\sigma'^{*},\rho^{*}}
    \prod_{i\in T^c}(z_{\sigma'(i),i}z^*_{\rho'(i),i}z^*_{\sigma'^{*}(i),i}z_{\rho'^{*}(i),i}) \\ 
    &=(N-k)!(!(N-k)),
\end{align}
where we used the fact that $\sigma=\sigma'^{*}$ and $\rho=\rho'^{*}$ is necessary to be nonzero.
Thus, the number of choices of $T$ and $T'$ has $\binom{N}{k}^2$ and the number of choices of $\sigma'$ and $\rho'$ is $(N-k)!(!(N-k))$ and we obtain
\begin{align}
    \mathbb{E}_Z[|f^{=(N-k)}|^2]
    &=\binom{N}{k}^2(N-k)!(!(N-k))\sum_{j=0}^k\binom{k}{j}^2j!(k-j)!(!(k-j))2^j.
\end{align}

\iffalse
As a remark, the square of the number of perfect matchings is
\begin{align}
    \sum_{k=0}^N\sum_{T,T'\subset [N],|T|=|T'|=k}\sum_{\sigma,\sigma',\rho'}1
    =\sum_{k=0}^N \binom{N}{k}^2k!(N-k)!(!(N-k))
    =(N!)^2,
\end{align}
where the following identity is used
\begin{align}
    \sum_{j=0}^k\sum_{i=0}^{k-j}\frac{(-1)^i}{i!j!}=1.
\end{align}
\fi

We now further upper bound the two-norm.
Here, we first use
\begin{align}
    \sum_{j=0}^k\binom{k}{j}^2j!(k-j)!(!(k-j))2^j
    &\leq \sum_{j=0}^k\binom{k}{j}^2j!((k-j)!)^22^j \\ 
    &=\sum_{j=0}^k\binom{k}{j}k!(k-j)!2^j \\ 
    &=e^2 k!\Gamma(k+1,2) \\ 
    &\leq e^2 k!\Gamma(k+1) \\ 
    &=e^2 (k!)^2.
\end{align}

Thus,
\begin{align}
    \mathbb{E}_Z[|f^{=(N-k)}|^2]
    &=\binom{N}{k}^2(N-k)!(!(N-k))\sum_{j=0}^k\binom{k}{j}^2j!(k-j)!(!(k-j))2^j \\ 
    &\leq e^2\binom{N}{k}^2(N-k)!(!(N-k)) (k!)^2 \\ 
    &\leq e^2\binom{N}{k}^2((N-k)!)^2 (k!)^2 \\ 
    &\leq e^2 (N!)^2.
\end{align}

\end{document}